\documentclass[aps,prl,graphicx,reprint,amssymb,citeautoscript,showpacs,superscriptaddress,floatfix
]{revtex4-1}

\newcommand{\rem}[1]{}

\usepackage[normalem]{ulem}
\usepackage[T1]{fontenc}
\usepackage{amsmath,amsfonts}
\usepackage{amssymb,textcomp}
\usepackage[xcdraw]{xcolor}
\usepackage{tikz,array}
\usepackage{bm}
\usepackage{graphicx, wrapfig}
\usepackage[caption=false]{subfig}
\usepackage{hyperref}
\usepackage{natbib}
\usepackage{comment}

\newcommand\beq{\begin{equation}}      
\newcommand\beqnn{\begin{eqnarray*}}   
\newcommand\beqa{\begin{eqnarray}}     
\newcommand\beqann{\begin{eqnarray*}}  

\newcommand\eeq{\end{equation}}        
\newcommand\eeqnn{\end{eqnarray*}}     
\newcommand\eeqa{\end{eqnarray}}       
\newcommand\eeqann{\end{eqnarray*}}    

\newcommand{\ave}[1]{\langle #1 \rangle}
\def\nl {\nonumber \\}

\newcommand\bi{\begin{itemize}}
\newcommand\ei{\end{itemize}}

\def\nl {\nonumber \\}

\newcommand{\eref}[1]{(\ref{#1})}

\newcommand{\fref}[1]{Fig.~\ref{#1}}

\newcommand{\Eref}[1]{Eq.~(\ref{#1})}

\newcommand{\al}[1]{\begin{align} #1 \end{align}}
\def\l0{L}

\begin{document}

\title{
Transient Casimir forces from quenches in thermal and active matter
} 

\author{Christian M. Rohwer}
\email[]{crohwer@is.mpg.de}

\affiliation{Max Planck Institute for Intelligent Systems, Heisenbergstr. 3, 70569 Stuttgart, Germany}
\affiliation{4th Institute for Theoretical Physics, Universit\"at Stuttgart, Pfaffenwaldring 57, 70569 Stuttgart, Germany}

\author{Mehran Kardar}
\affiliation{Department of Physics, Massachusetts Institute of Technology, Cambridge, Massachusetts 02139, USA}

\author{Matthias Kr\"uger}
\affiliation{Max Planck Institute for Intelligent Systems, Heisenbergstr. 3, 70569 Stuttgart, Germany}
\affiliation{4th Institute for Theoretical Physics, Universit\"at Stuttgart, Pfaffenwaldring 57, 70569 Stuttgart, Germany}

\date{\today}

\begin{abstract}
We compute fluctuation-induced (Casimir) forces for classical systems after a temperature quench. Using a generic coarse-grained model for fluctuations of a conserved density, we find that transient forces arise even if the initial and final states are force-free. 
In setups reminiscent of Casimir (planar walls) and van der Waals (small inclusions) interactions,
we find comparable exact universal expressions for the force.
Dynamical details only scale the time axis of transient force curves. 
We propose that such quenches can be achieved, for instance, in experiments on active matter,
employing tunable activity or interaction protocols.

\end{abstract}

\maketitle 

Fluctuation-induced forces~\cite{kardargolestanian1999} are well-known for quantum electromagnetic fields~\cite{casimir1948,bordag}, as well as a host of classical systems~\cite{fisherdegennes,hertlein2008,gambassiCCFPRE2009,garciachan2002,garciachan2006,fukutowettingfilms2005,linzandi2011}. 
At thermal equilibrium with an infinite correlation length in the medium (e.g. near a critical point~\cite{fisherdegennes}), the force $F$ per area $A$ between two parallel plates at a distance $L$ in $d=3$ dimensions takes the general form 
\begin{align}
\frac{F}{A}\propto \frac{k_BT}{L^3}\,,\label{eq:1}
\end{align}
in the classical limit (with Boltzmann's constant $k_B$ and temperature $T$). Prefactors, which may depend on boundary conditions, are typically of order unity. 


Various non-equilibrium aspects of fluctuation-induced forces in electromagnetic~\cite{Antezza08,krueger2011} and 
thermal systems~\cite{gambassi2006,gambassi2008EPJB,deangopinathan2009JStatMech,deangopinathan2010PRE,gambassi2013prl,deanpodgornik2014} have been explored. 
Particularly intriguing are {\it long-ranged} forces appearing in non-equilibrium situations
where a corresponding force is absent in equilibrium.
These arise due to dynamic conservation laws, which generally produce long-ranged 
correlations out of equilibrium~\cite{mukamelkafri1998}. 
Examples include fluids subject to gradients in temperature~\cite{kirkpatricksengers2013},
particles diffusing in a density gradient~\cite{aminovkardarkafri2015}, 
and driven systems~\cite{wadasasa2003,sotogranular2006,shaebaniwolf2012}. 
The corresponding non-equilibrium forces are generally non-universal and depend on
 dynamical details. 

In this paper, we consider the stochastic dynamics of a conserved field (density) in systems exhibiting short-ranged (local) correlations in steady state. We find that changing the noise strength results in transient forces at intermediate times, described by the universal, detail-independent form of Eq.~\eqref{eq:1} (for parallel surfaces), but with a time-dependent amplitude. 
The latter is governed by a time scale set by the diffusivity of the field. For the case of small objects immersed in the medium, the force resembles classical (equilibrium) van der Waals interactions. Correlations again become short-ranged at long times, and long-ranged forces decay with power laws in time. The model is mathematically equivalent to the well-known ``model B''~\cite{hohenberg,kardarbook,onukibook} dynamics. Therefore we shall refer to the strength of the noise as temperature, and to the protocol for changing the strength of noise as a quench. Since this description arises naturally upon coarse-graining generic systems without long-ranged correlations, we expect the results to be observable in a variety of setups. Notably, these include certain cases of active matter, as discussed later in the manuscript.

Consider a classical system (e.g. a fluid in equilibrium, 
shaken granular matter, or active particles) 
characterized by a local density $c(\bm x)$, fluctuating around an average 
value of $\ave{c (\bm x)}$ in $d$ spatial dimensions, $\bm x \in \mathbb R^d$. 
For a {\it conserved} density the fluctuating field  
$\phi(\bm x) = c(\bm x) - \ave{c (\bm x)}$ is constrained to evolve according to
$\partial_t \phi(\bm x,t) = - \nabla\cdot {\bf j}$.
The current ${\bf j}$ is comprised of a deterministic component ${\bf j}_d$, and a stochastic
component ${\bf j}_s$. The former originates from the interactions amongst the 
microscopic constituents (including any obstacles), the latter from thermal fluctuations
or random changes in active driving forces.  Both contributions can be complex at the microscopic level. However, for short-ranged interactions, simplified descriptions can be obtained by coarse-graining beyond relevant length scales, e.g. the correlation length for fluids, or the so-called run length for active Brownian particles. Symmetry considerations then restrict the deterministic current to 
\al{
{\bf j}_d=-\mu\nabla [m(\bm x) \phi(\bm x,t)+\cdots] ,
\label{eq:Jd}
}
where we have allowed for a non-uniform ``mass'' $m(\bm x)$ to later account for 
inclusions and boundaries in the field. Higher order terms in $\phi$ and $\nabla\phi$
can be added, but will be irrelevant at large length scales (as seen e.g. from dimensional analysis~\cite{kardarbook}). This leads to the stochastic diffusion equation~\cite{kardarbook},  
\al{
\partial_t \phi(\bm x,t) &=   \mu \nabla^2 [m(\bm x) \phi(\bm x,t)] + \eta(\bm x,t)\,.
\label{eq:langevin}
}
The noise has zero mean, and its contribution to the above equation has covariance
\al{
\ave{\eta(\bm x,t)\eta(\bm x',t')} &= - 2D  \nabla^2 \delta^d (\bm x - \bm x')\delta(t-t').
\label{eq:noise}
}

Equations~\eqref{eq:langevin} and \eqref{eq:noise} are equivalent to ``model B''
dynamics of a field subject to a local Hamiltonian~\cite{onukibook,kardarbook}, 
$H[\phi] = \int \mathrm d^d x\;  \frac {m(\bm x)} 2\phi^2(\bm x)$, and
a ``mobility'' $\mu$.
This equivalence indicates that the correlation functions in steady state
satisfy $\ave{\phi(\bm x)\phi(\bm x')}=\delta^d(\bm x-\bm x')\frac{D}{\mu m}$.
This corresponds to the equilibrium ensemble with the Boltzmann distribution, 
$P_{eq} \propto \exp(-H [\phi]/k_B T)$, provided that the noise satisfies the
Einstein relation $D=\mu k_BT$. 
The mass $m$ is thus a measure of (in)compressibility of the density field.
For active systems, we can adopt the same notation, but with an
effective temperature. For example, active Brownian particles (coarse-grained
beyond the run length) can in many aspects be described by  Eqs.~\eref{eq:langevin} and \eref{eq:noise} with an effective temperature~\cite{loiEffectiveT2008},
related to the self-propulsion velocity, that can be orders of magnitude larger than
room temperature. We therefore assume that in these cases forces are found equivalently (see below).

Long-ranged fluctuation-induced interactions occur if inclusions disrupt
long-ranged correlations in the medium~\cite{kardargolestanian1999}. Since correlations of the field $\phi$
in steady state are local, no long-ranged steady state Casimir forces are expected.
We investigate what happens if the stochastic force is suddenly changed,
specifically in a quench at time $t=0$ from an initial state with $\phi=0$~\cite{deangopinathan2010PRE}, to a finite `temperature' $T$ (or finite $D$). 

\begin{figure}[t]
 \begin{center}
 \includegraphics[width=\columnwidth]{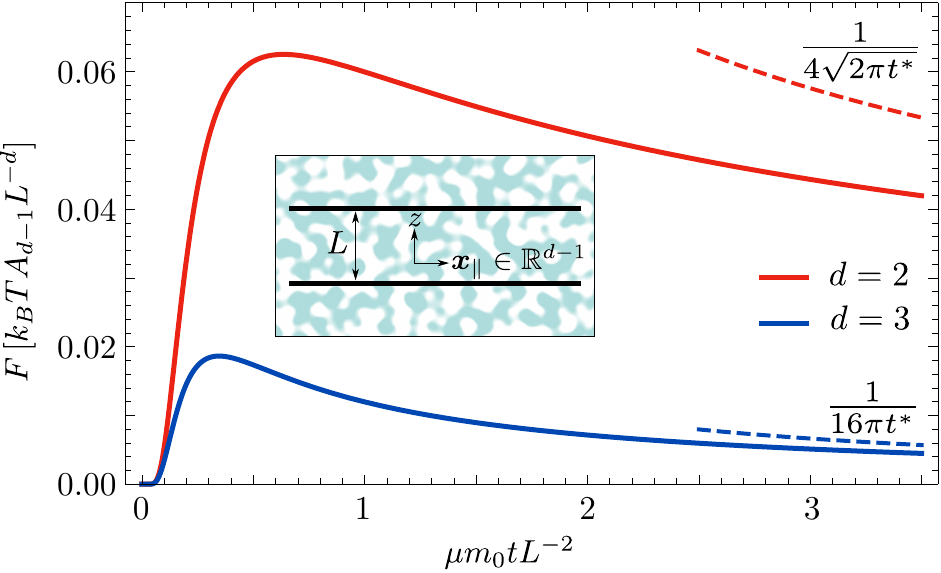}
 \end{center}
 \vspace{-0.6cm}
\caption{The transient force $F$ between two parallel plates from \Eref{eq:ForceP}, as a function
of  rescaled time $t^*=\mu m_0 tL^{-2}$. The dotted lines indicate the long-time asymptotes. 
The force is always attractive.}
 \label{fig:FPlates} 
\end{figure}

Consider first two parallel, impenetrable plates (as in the inset of Fig.~\ref{fig:FPlates}) inserted in a medium with uniform 
mass $m_0$. Impenetrability gives rise to no-flux boundary conditions, so that the normal component of the 
total current ${\bf j}$ vanishes at all times on the surfaces of the plates, situated at $z=0$ and $z=L$~\cite{diehl1994,diehlwichmann1995}. This is guaranteed by constraining $\phi$ to Neumann modes 
$\phi_n \propto \cos(k_{n} z)$ with $k_n = n\pi/L$, $n=0,1,\cdots$, for $z$ normal to the surfaces. 
(Coordinates parallel to the surface are denoted as $\bm x_\parallel \in \mathbb R^{d-1}$.) 
A similar decomposition also ensures the no-flux condition for the stochastic current. 
A straightforward computation yields the time-dependent (transient) field correlations for $z$ and $z'$ between $0$ and $L$,
\al{
\ave{\phi(z,t)\phi(z',t)}&= \frac{2k_B T} {m_0L} {\sum_{n=0}^\infty} \mathcal N_n \cos(k_{n} z)\cos(k_{n} z')  \nl
&\qquad\times \int {\mathrm d\bm p}\; \frac{1-e^{-2(k_n^2 +p^2) \mu m_0 t}}{(2\pi)^{d-1}}\,,
\label{eq:phi}
}
where $\bm p\in\mathbb R^{d-1}$ is the Fourier vector conjugate to $\bm x_\parallel$, and 
$\mathcal N_n = 1-\frac 1 2\delta_{n,0}$. 
For fundamental fields such as electromagnetism, forces can be obtained from the stress tensor.
Applicability of such a procedure to coarse-grained fields describing a system out of equilibrium is debatable,
but Ref.~\cite{deangopinathan2010PRE} provides a general and powerful scheme, applied to systems with infinite equilibrium correlation lengths, subject to non-conserved dynamics. 
This scheme is not directly applicable here, since we use (no-flux) boundary conditions, as opposed to introducing 
terms in the Hamiltonian that mimic the obstacles.
We use {\it actio et reactio}, and find the force acting on the medium instead. 
Introducing an external pseudo-field which balances the deterministic current ${\bf j}_d$ \cite{deanlecture} in Eq.~\eqref{eq:langevin}, we find the force density ${\bf f}$ acting on the field, ${\bf f}=\phi{\bf j}_d/\mu$. The force per area $A_{d-1}$ acting on the wall at $z=0$ (which is minus the force acting on the medium) then reads
\al{
\frac{
F(t)}{A_{d-1}} =- \frac {m_0} 2\big[\ave{\phi^2(z=0,t)}-\ave{\phi^2(z=0,t)}_{L\to\infty}\big].
\label{eq:FfromC}
}
The second term, evaluated for $L\to\infty$, represents the pressure on the plate from the medium outside the cavity. 
This way of computing forces agrees with the non-equilibrium stress tensor found in Ref.~\cite{deanlecture}. 
\rem{One may notice that the force in Eq.~\eqref{eq:FfromC} is exactly {\it opposite} to the force that would be found from 
naive use of the equilibrium form of the stress tensor. }
Note that the equilibrium force is exactly zero as the corresponding correlation function is independent of $L$,
and cancels out  in Eq.~\eqref{eq:FfromC}.

The time-dependent correlation function in Eq.~\eqref{eq:phi} exhibits a trivial divergence for all times $t>0$, 
corresponding to the $\delta$-function form of the local correlation functions. However, this divergence does not 
contribute to the net force, and is removed when taking the difference in Eq.~\eqref{eq:FfromC}. 
The resulting force takes a universal form in terms of the rescaled time $t^* = \mu m_0 t/L^2$, 
\al{
\frac{
F(t)
}{A_{d-1}}= \frac{k_B T}{L^d} \frac{\vartheta _3\left(0,e^{-\frac{1}{2 t^*}}\right)-1}{ (8\pi t^*)^{d/2}},
\label{eq:ForceP}
}
where $\vartheta_3(0,q) = 1+2\sum_{n=1}^\infty q^{n^2}$ is the Jacobi elliptic function of the third kind~\cite{abramowitzstegun}. This is our first main result. The force in Eq.~\eqref{eq:ForceP} is the product of two factors. The second, time-dependent factor is free of units, and is shown in~\fref{fig:FPlates} for $d=2$ and $d=3$~\footnote{Note that in $d=1$ the force
does not decay to zero at large times. This is because the single mode with $n=0$, which encodes the conservation
law, makes a finite contribution to the overall result.}. 
In both the $\phi=0$ state before the quench ($t<0$), as well as for asymptotically long times after the quench ($t\to\infty$), 
the force vanishes due to the locality of the correlations. 
However, there is a finite and attractive transient force at intermediate times.
 At short times, the series for $\vartheta _3$ indicates a leading essential singularity $F\sim e^{-\frac{1}{2 t^*}}/(t^*)^{d/2}$. 
The force reaches a maximum  at $t^*\approx \frac{1}{2}$, in a time scale set by diffusion across the separation $L$~\cite{deangopinathan2010PRE}. For long times $ \vartheta _3$  grows as $\sqrt{t^*}$, leading to a power law decay
of the force as $(t^*)^{-(d-1)/2}$. This scale-free decay is associated with relaxation of the unbounded modes (parallel to the surfaces as well as in the semi-infinite system faced by the outside surfaces). 

The overall amplitude in \Eref{eq:ForceP} has the form of Eq.~\eqref{eq:1},
which describes equilibrium forces in scale free media. Thus, strikingly, and in contrast to previously found non-equilibrium 
Casimir forces, the amplitude of the force is independent of (dynamical) details of the system. 
These enter only in the scaling of $t^*$ so that, for different systems, merely the time axis is rescaled. 
Specific comparison to the equilibrium force of a Gaussian critical theory with Hamiltonian  $H[\phi]= \int \mathrm d^d x\;\frac{r}{2}  (\nabla\phi)^2$ (and Dirichlet boundary conditions) shows that the maximum of the transient force computed 
here is very similar, smaller by a factor of $0.39$ ($d=2$) and $0.48$ ($d=3$). 
Since equilibrium fluctuation-induced forces have been measured in various systems (see e.g. Ref.~\cite{hertlein2008}), 
we expect the transient forces to be measurable as well (even when not accounting for the much larger 
expected effective temperatures).


As a second example, consider the transient force  between two small inclusions of volumes $V_1$ and $V_2$ 
embedded in the medium, modeled by ``compressibilities'' deviating by $c_i$ from the surrounding 
medium, i.e., 
\al{
m(\bm x)=
\begin{cases}
 m_0,\quad &\bm x \notin V_i,\\
 m_0+c_i,\quad &\bm x \in V_i.
\end{cases}
}
This case is well-suited to the formalism developed in Ref.~\cite{deangopinathan2010PRE}, where the force is found from $F(t)=-\langle\frac{\partial H(t)}{\partial L}\rangle$, with $L$ being the center-to-center distance of the objects.  
Solving the linear Langevin Eq.~\eqref{eq:langevin} in Laplace space~\footnote{
The Laplace transform is {$\mathcal L f(s)=\int_0^\infty \mathrm dt\; e^{-s t} f(t)$}.
}, treating the external bodies as
perturbations to the bulk, an intriguing connection is observed~\cite{deangopinathan2010PRE}: just as  
in equilibrium, the force between embedded objects is related to the bulk correlation function of $\phi$. 
However, in the non-equilibrium case, this relation is formulated in Laplace space. 
Specifically, in the limit where the distance $L$ is large compared to the extension of the objects, $L\gg V_i^{1/d}$, 
the Laplace transformed average force is~\cite{deangopinathan2010PRE}
\al{
\mathcal L  F(s)&= k_B T \frac{c_1 V_1}{m_0}\frac{ c_2 V_2}{m_0} \frac{1}{s} \;G_s(L)\partial_LG_s(L).
\label{eq:F(s)}
}
Here $G_s(X) = \mathcal L [\ave{\phi(0,t)\phi(X,t)}](s)$ is the Laplace transform of the transient (post-quench) equal-time correlation function for two points separated by a distance $X$ in the medium. %
\begin{figure}[t]
 \begin{center}
 \includegraphics[width=\columnwidth]{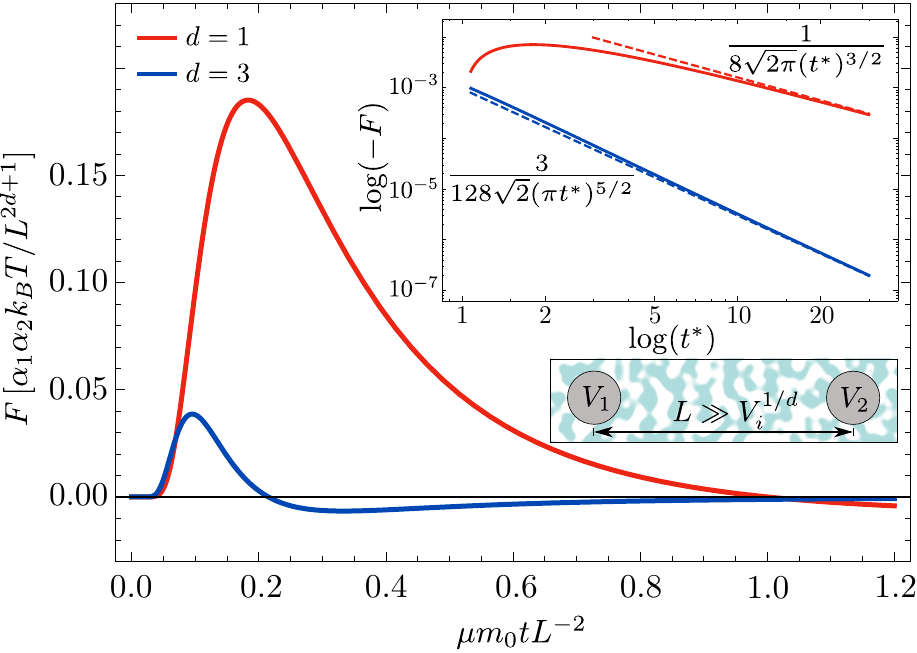}
 \end{center}
 \vspace{-0.6cm}
\caption{The force between two small embedded objects (\Eref{eq:Finc}) as a function of dimensionless time $t^* =\mu m_0 tL^{-2}$. Dotted lines in the insets are long-time asymptotes as given.}
 \label{fig:test2} 	
\end{figure}
The correlation functions in $d=1$ and $d=3$ have the explicit forms (note that this is the same quantity as in \eref{eq:phi} but for $L\to\infty$)
\begin{align}
\ave{\phi(0,t)\phi(X,t)} 
&=\begin{cases}
-\frac{k_B T }{m_0 X }\frac{e^{-\frac{1}{8 t^*}}}{(8 \pi  t^*)^{1/2}},&d=1,\\
-\frac{k_B T }{m_0 X^3 }\frac{e^{-\frac{1}{8 t^*}}}{(8 \pi  t^*)^{3/2}},&d=3,
\end{cases}
\label{eq:bulkC(t)}
\end{align}
with $t^* = \mu m_0 t/X^2$. These expressions are  quite similar to the force between two surfaces in Eq.~\eqref{eq:ForceP}. 
Indeed, the nature of this correlation function is the origin of the transient forces: starting from zero, it approaches a maximum, set by the time necessary for diffusion across the distance $X$, and decays with a power law for $t\to\infty$. Its Laplace transform, $G_s(X)\propto e^{-X\sqrt{s/2m_0\mu}}$, is also illustrative, with $\sqrt{s/2m_0\mu}$ playing the role of an inverse correlation length. For small $s$ (large times), correlations are long-ranged, while for large $s$ (short times) they are exponentially cut off, since distant points have not yet communicated via diffusion. Technically, the long-ranged character of the transient correlation function enters via the inverse of the Laplacian in Eqs.~\eqref{eq:langevin} and \eqref{eq:noise}; e.g., in $d=3$, $[\nabla^2\delta^3(\bm x- \bm x')]^{-1}=-\frac{1}{4\pi|\bm x- \bm x'|}$.

Laplace inversion of \Eref{eq:F(s)} yields the transient force, which reads (we\rem{ exclude the case of $d=2$, and} define $\alpha_i = c_i V_i / m_0$)
\begin{align}\label{eq:Finc}
F(t)&= k_B T \frac{\alpha_1\alpha_2}{L^{2d+1} } \Xi_d(t^*)\,,
\end{align}
with Eq.~\eqref{eq:bulkC(t)} leading to the time-dependent amplitude
\begin{align}
\Xi_d(t^*) 
&=e^{-\frac{1}{2 t^*}}\times
\begin{cases}
\frac{(1-t^*)}{16\sqrt{2\pi}\left( t^*\right)^{5/2}},&d=1, \\
\frac{[1-t^*(3t^* + 4)]}{256\sqrt 2 \pi^{5/2}\left( t^*\right)^{9/2}},&d=3.
\end{cases}
\end{align}
As in the case of parallel surfaces,  the force between  small objects (depicted in \fref{fig:test2})
rises from zero and reaches a maximum (here at $t^*\approx 0.15$).
However, in sharp contrast to the former, the force changes  sign from attractive at short times to repulsive at long times 
at around $t^*\approx 1$. 
This can be interpreted as the effect of back-flow from the diffusive front of fluctuations which passes beyond the
inclusions  at $t^*\approx 1$. 
In the case of parallel plates, fluctuations are confined and cannot pass beyond the obstacles. 
The long-time decay of the force is a manifestation of the well-known long-time correlations in
conserved dynamics (cf. Eq.~\eref{eq:bulkC(t)}).

Just as in the case of parallel plates, the overall amplitude of the force between inclusions is independent of 
dynamical details, which only scale the time axis. Furthermore, the force in \Eref{eq:Finc} resembles the 
van der Waals force between two particles with polarizabilities $\alpha_i$ in the classical limit~\cite{boyer1975}. 
Hence, as the force in Eq.~\eqref{eq:ForceP}, it acquires a very well-known and studied form. 
The notation $\alpha_i = c_i V_i / m_0$ is also motivated by this analogy, and just as in the case
of (electromagnetically) polarizable particles, $\alpha_i$ is proportional to the particle's volume and 
(optical or compressibility) contrast~\cite{Tsang}. 
Finally, we comment that this analogy carries a practical message: 
$\alpha_i$ in \Eref{eq:Finc} is related to the perturbative solution of Eq.~\eqref{eq:langevin} for 
a single object $i$ in isolation. It can thus be measured independently in a ``scattering'' experiment, 
so that $\alpha_i$ is not a free fit parameter when applying \Eref{eq:Finc} in a given experiment.


The temperature quench investigated here can be realized experimentally in various ways. In addition to directly changing temperature, there are various experimental techniques to rapidly change interparticle potentials. 
Such changes (e.g. from hard to soft) in compressibility have the same effect as a temperature quench: 
initially, fluctuations are suppressed ($\ave{\phi\phi}\approx0$), and then suddenly start growing, 
giving rise to the phenomena analyzed here. Examples of tunable interparticle potentials include 
thermosensitive particles whose radii change strongly over a very small temperature range~\cite{ballauff2006}, 
or magnetic nano-colloids whose interactions can be strongly tuned with an externally applied magnetic field~\cite{maretkeim2004}.

A particularly timely class of experimental candidates concerns the aforementioned {\it active matter} systems with effective temperatures~\cite{loiEffectiveT2008}. 
Importantly, activity can often be tuned externally, for instance for Brownian particles with tunable illumination-induced activity~\cite{buttinoni2012active}, or agitated granular beads~\cite{kudrollikantorkardar2009,kumarramaswamisood2014}, so that quenches can be applied easily. It is also relevant that effective temperatures of such systems are often much larger than experimental (room) temperatures. This acts in favour of the forces in Eqs.~\eqref{eq:ForceP} and \eqref{eq:Finc}, which are proportional to the (effective) temperature.

To conclude, classical systems with a conserved density undergoing temperature quenches (or changes in noise or activity) show transient Casimir forces with universal amplitudes, analogous to equilibrium forces in scale free media. Dynamical details scale the time axis of the forces, which are maximal at a time corresponding to diffusion across  distances between obstacles. 
The transient forces depend on the history of quenching. Therefore it may be possible to generate persistent non-equilibrium forces through periodically varying temperature protocols; this will be addressed in future work. 
The methods presented here can be adapted to various geometries and a broad class of non-equilibrium systems.


\begin{acknowledgments}
We thank D.~S.~Dean, G.~Bimonte, T.~Emig, N.~Graham,  R.~L.~Jaffe and M.~F.~Maghrebi for discussions. This work was supported by MIT-Germany Seed Fund Grant No.~2746830.  Ma.Kr.~and C.M.R.~are supported by Deutsche Forschungsgemeinschaft (DFG) Grant No.~KR 3844/2-1. M.Ka.~is supported by the NSF through Grant No. DMR-12-06323.
\end{acknowledgments}

\bibliographystyle{apsrev4-1}
\bibliography{references.bib}

\end{document}